\documentclass{elsart}
\usepackage{amsfonts,amsmath,amssymb,array,graphicx,tabularx}
\usepackage[mathscr]{eucal}

\begin{document}

\begin{frontmatter}

\title{On the parabolic equation method in internal wave propagation\thanksref{GOV}}

\author{M.Yu.~Trofimov}\ead{trofimov@poi.dvo.ru},
\author{S.B.~Kozitskiy}\ead{skozi@poi.dvo.ru}~and \author{A.D.~Zakharenko}\ead{zakharenko@poi.dvo.ru}

\address{Il'ichev Pacific oceanological institute, Baltiyskaya St. 43, Vladivostok, 41, 690041, Russia }
\thanks[GOV]{This work is supported by the Program No. 14 (part 2) of the Presidium of
the Russian Academy of Science.}

\journal{Ocean Modelling}\volume{17}\issue{4}
\setcounter{firstpage}{327} \setcounter{lastpage}{337}
\pubyear{2007}

\begin{abstract}
A parabolic equation for the propagation of periodic internal waves
over varying bottom topography is derived using the multiple-scale
perturbation method. Some computational aspects of the numerical
implementation are discussed. The results of numerical experiments
on propagation of an incident plane wave over a circular-type shoal
are presented in comparison with the analytical result, based on
Born approximation.
\end{abstract}

\end{frontmatter}

\section{Introduction}
\label{section1}

The parabolic equation method, as an approximation to some elliptic
problems, has been extensively used in mathematical physics. It was
introduced by Leontovich \& Fock \cite{foc} in the theory of
electromagnetic wave propagation, applied in monograph of Babich \&
Buldyrev \cite{bab} to various diffraction problems and developed by
Tappert \cite{tap} and his successors in underwater acoustics. Later
it appeared also in the theory of surface water wave propagation in
works of Liu \& Mei \cite{liu}, Radder \cite{rad}, Kirby \&
Dalrymple \cite{kir} and others.
Since then, it has been proven through the contribution of several
authors to be an effective model for dealing rapidly and accurately
with propagation problems in coastal areas.
\par
The standard, or narrow angle parabolic wave equation has the form
of the quantum mechanical non-stationary Schr\"odinger equation and
describes the propagation of waves in weakly inhomogeneous media, at
small angles with a preferred direction (taken to be in this paper
the $x$-direction). Being of the evolution type equation, it can be
solved in the frame of the initial-boundary value problems and has
an advantage over the geometric optic method in describing the wave
field near a caustic. In the theory of the surface water wave
propagation the parabolic equation method now can be considered as
classic and is well explained in Mei's monograph \cite{mei}. In this theory
some more complicated models, which
describe wide-angle propagation \cite{dal-kir} and include some
nonlinear effects \cite{kir,liu-yoon}, were also developed.
\par
In the theory of internal wave propagation, though the geometric optic method
is well-developed \cite{kel-mow,mir,vor}, little is known about the parabolic equation method,
except the Kadomtsev-Petviashvili equation, which in the variable topography case was derived
for the  interfacial waves in \cite{chen-liu}.
\par
The case of continuous stratification is in close analogy with the acoustic case,
where the adiabatic mode parabolic equations were first derived by the factorization method in
\cite{collins} and later by the multiple-scale method in
\cite{trof1} and \cite{trof2}.

\par
The aim of our work is to obtain a standard parabolic equation for the
internal wave propagation in the most simple cases and briefly
discuss related computational aspects. Rotation, which is ignored at this
stage, can be included in the model as weak rotation \cite{chen-liu},
and will be considered in our further publications.

As an illustration we present some numerical calculations performed
in the case of refraction of an incident plane wave on a
circular-like shoal.

\section{Formulation and scaling}
\label{section2} The system of linear equations, describing small
amplitude motions of stratified inviscid fluid with harmonic
dependence on time $t$ by the factor $e^{-\mathrm{i}\omega t}$, from
which we are starting, is
\begin{equation}\label{base}
\begin{split}
-\mathrm{i}\omega\rho_0u+P_x=0  \\
-\mathrm{i}\omega\rho_0v+P_y=0  \\
-\mathrm{i}\omega\rho_0w+P_z+g\rho_1=0 \\
-\mathrm{i}\omega\rho_1+w\rho_{0z}=0  \\
u_x+v_y+w_z=0
\end{split}
\end{equation}
Here $x$, $y$, and $z$ are the Cartesian coordinates (vertical axis is
directed upward),
$\rho_0=\rho_0(z)$ is the undisturbed density, $P$ is the pressure,
$g$ is the gravity acceleration, $\rho_1=\rho_1(x,y,z)$ is the perturbation
of density due to motion, and $u$, $v$ and $w$ are respectively  the $x$, $y$ and $z$
components of velocity.
\par
We consider these equations with the boundary conditions
\begin{equation} \label{bc}
\begin{split}
 w =  0 \quad  &\mbox{at} \quad z=0 \\
 w+uH_x+vH_y=0 \quad  &\mbox{at} \quad  z=-H
\end{split}
\end{equation}
where $H=H(x,y)$ is the bottom depth.
\par
To begin the multiple-scale procedure \cite{nay}, we introduce a
small parameter $\epsilon$, the slow variables $X=\epsilon x$ and
$Y=\epsilon^{1/2}y$, the fast variable $\xi=(1/\epsilon)\Theta(X,Y)$
and expand the dependent variables as follows:
$$
u  =  u_0+\epsilon u_1+\ldots \mbox{,\qquad}
v  =  \epsilon^{1/2}v_{1/2}+\ldots \mbox{,}
$$
$$
w  =  w_0+\epsilon w_1+\ldots \mbox{,\qquad}
P  = P_0+\epsilon P_1+\ldots
$$
The scaling used in the definition of the slow variables is characteristic
for the parabolic equation method \cite[Section 4.10]{foc,mei}, with $x$-direction
as the principal propagation direction and $y$-direction as the transverse direction.
\par
From now on we assume that
$$
H  =  H_0(X)+\epsilon H_1(X,Y)\,,
$$
so  $H_0$ is independent of $Y$.

\par
We also expand $\rho_1$ in powers of $\epsilon$
$$
\rho_1=\rho_{10}+\epsilon\rho_{11}+\ldots\,.
$$
\par
Substitution of these expansions into system  (\ref{base}) and the
boundary conditions (\ref{bc})
 and changing the partial derivatives for the prolonged ones by the rules
\begin{equation*}
\begin{split}
&\frac{\partial}{\partial x}  \rightarrow \epsilon\frac{\partial}{\partial
X}+\Theta_X\frac{\partial}{\partial
\xi}\,, \\
&\frac{\partial}{\partial y}  \rightarrow
\epsilon^{1/2}\frac{\partial}{\partial Y}+
\Theta_Y\frac{\partial}{\partial \xi}\,,
\end{split}
\end{equation*}
leads to the system of equations
\begin{align}
%\begin{equation} \label{eq7}
 -\mathrm{i}\omega\rho_0(u_0+\epsilon u_1+\ldots)+\epsilon
(\frac{\partial}{\partial
X}+\frac{1}{\epsilon}\Theta_X\frac{\partial}{\partial \xi})
 (P_0+\epsilon P_1+\ldots)=0&\,,\label{eq7}\\
%\end{equation}
%\begin{equation} \label{eq8}
 -\mathrm{i}\omega\rho_0(\epsilon^{1/2}v_{1/2}+\ldots)+
\epsilon^{1/2}(\frac{\partial}{\partial
Y}+\frac{1}{\epsilon}\Theta_Y\frac{\partial}{\partial \xi}) \label{eq8}
(P_0+\epsilon P_1+\ldots)=0&\,,\\
%\end{equation}
%\end{align}
%\begin{equation} \label{eq9}
-\mathrm{i}\omega\rho_0(w_0+\epsilon w_1+\ldots)+ P_{0z}+\epsilon
P_{1z}+\ldots+g(\rho_{10}+\epsilon\rho_{11}+\ldots)=0&\,,\label{eq9}\\
%\end{equation}
%\begin{equation} \label{eq10}
-\mathrm{i}\omega(\rho_{10}+\epsilon\rho_{11}+\ldots)+ (w_0+\epsilon
w_1+\ldots)\rho_{0z}=0&\,,\label{eq10}
%\end{equation}
\end{align}
\begin{equation}\label{eq11}
\begin{split}
\epsilon(\frac{\partial}{\partial
X}+\frac{1}{\epsilon}&\Theta_X\frac{\partial}{\partial \xi})
(u_0+\epsilon u_1+\ldots)\\
+ &\epsilon^{1/2}(\frac{\partial}{\partial Y}+\frac{1}{\epsilon}\Theta_Y\frac{\partial}{\partial
\xi})(\epsilon^{1/2}v_{1/2}+\ldots)
 +w_{0z}+\epsilon w_{1z}+\ldots=0\,,
\end{split}
\end{equation}
with the boundary conditions
\begin{equation}\label{eq12}
 w_0+\epsilon w_1+\ldots =  0 \qquad \text{at}\quad z=0\,,
\end{equation}
\begin{equation}\label{eq13}
\begin{split}
 w_0+&\epsilon w_1+\ldots + \epsilon(u_0+\epsilon u_1+\ldots)
\frac{\partial}{\partial X}(H_0 + \epsilon H_1)  \\
 & +\epsilon^{1/2}(\epsilon^{1/2}v_{1/2}+\ldots)
\frac{\partial}{\partial Y}(H_0+\epsilon H_1)=0
  \qquad \text{at}\quad  z=-(H_0+\epsilon H_1)\,.
\end{split}
\end{equation}

\section{The parabolic equation}

Equating coefficients in Eqs.~(\ref{eq7}-\ref{eq13}) of like powers
of $\epsilon$, we obtain equations describing $u_l, w_l, v_{l+1/2}$,
$l=0,1,\ldots$.
\par
At the order $-1/2$ in $\epsilon$ we obtain from (\ref{eq8})
$$
\Theta_Y P_{0\xi}=0\,,
$$
and put $\Theta_Y=0$, so in the sequel $\Theta$ depends only on $X$, $\Theta=\Theta(X)$.
\par
At the zeroth order in $\epsilon$ we have
\begin{equation} \label{ord13}
-\mathrm{i}\omega\rho_0u_0+\Theta_X P_{0\xi}=0\,,
\end{equation}
\begin{equation} \label{ord14}
-\mathrm{i}\omega\rho_0w_0+P_{0z}=-g\rho_{10}\,,
\end{equation}
\begin{equation} \label{ord15}
-\mathrm{i}\omega\rho_{10}+w_0\rho_{0z}=0\,,
\end{equation}
\begin{equation} \label{ord16}
\Theta_Xu_{0\xi}+w_{0z}=0\,,
\end{equation}
with the boundary conditions
\begin{equation*}
\begin{split}
w_0 =  0 \quad &\mbox{at} \quad z=0\,, \\ w_0 =  0 \quad &\mbox{at}
\quad z=-H_0\,.
\end{split}
\end{equation*}
\par
After substitution in Eq.~(\ref{ord14}) $\rho_{10}$ from
Eq.~(\ref{ord15}) we get
\begin{equation} \label{ord18}
(\omega^2\rho_0+g\rho_{0z})w_0+\mathrm{i}\omega P_{0z}=0\,.
\end{equation}
Twice differentiating this equation with respect to
$\xi$ and using the expression
\begin{equation*}\label{ord19}
P_{0z\xi\xi}= -\frac{1}{(\Theta_X)^2}\mathrm{i}\omega(\rho_0
w_{0z})_z,
\end{equation*}
obtained from Eqs.~(\ref{ord13}) and (\ref{ord16}), we get
\begin{equation*}\label{ord20}
(\Theta_X)^2(\omega^2\rho_0+g\rho_{0z})w_{0\xi\xi}+\omega^2(\rho_{0z}w_{0z})_z=0\,.
\end{equation*}
We seek a solution of this equation in the form of WKB-type anzatz
$w_0=A(X,Y)\phi(X,z)e^{i\xi}$, where $\phi$ is an eigenfunction with
the eigenvalue $k^2=(\Theta_X)^2$ of the spectral problem
\begin{equation} \label{ord21}
\begin{split}
&\omega^2(\rho_0\phi_z)_z-k^2(\omega^2\rho_0+g\rho_{0z})\phi=0\,,  \\
&\phi(0)=\phi(-H_0)=0\,,
\end{split}
\end{equation}
normalized by the condition
\begin{equation} \label{ord39}
\frac{\omega^2}{k^2}\int^0_{-H_0}\rho_0\cdot(\phi_z)^2dz=
-\int^0_{-H_0}(\omega^2\rho_0+g\rho_{0z})\phi^2dz=1\,.
\end{equation}
This problem is known as the main spectral problem of the linear internal wave theory.

At $O(\epsilon^{1/2})$ we have only one equation
\begin{equation}
-\mathrm{i}\omega\rho_0v_{1/2}+P_{0Y}=0\,, \label{ord27}
\end{equation}
which express the balance in the transverse direction for the quantities of order $<O(\epsilon)$.

The system of equations at the first order in $\epsilon$ is
\begin{equation}\label{ord22}
-\mathrm{i}\omega\rho_0u_1+\Theta_XP_{1\xi}+P_{0X}=0\,,
\end{equation}
\begin{equation}\label{ord23}
-\mathrm{i}\omega\rho_0w_1+P_{1z}=-g\rho_{11}\,,
\end{equation}
\begin{equation}\label{ord24}
u_{0X}+\Theta_Xu_{1\xi}+v_{1/2Y}+w_{1z}=0\,,
\end{equation}
\begin{equation}\label{ord25}
-\mathrm{i}\omega\rho_{11}+w_1\rho_{0z}=0\,,
\end{equation}
with the boundary conditions
\begin{equation}\label{ord25a}
w_1 = 0 \qquad \mbox{at} \qquad z=0\,,
\end{equation}
and
$$
w_0(z)+ \epsilon w_1(z)+\epsilon u_0(z) H_{0X} = 0 \qquad
\mbox{at}\qquad z=-H_0-\epsilon H_1\,.
$$
Expanding velocities in Taylor series with respect to $z$ at
$z=-H_0$ and collecting terms at $\epsilon^1$, we reduce the last
boundary condition on $z=-H_0$
\begin{equation}\label{ord26}
w_1 -w_{0z}H_1+u_0 H_{0X} = 0 \qquad \mbox{at}\qquad z=-H_0\,.
\end{equation}
\par
Considerations, similar to those in the derivation of the spectral
problem Eq.~(\ref{ord21}) (the details are given in
Appendix~\ref{A1}), lead at $O(\epsilon)$ to the equation
\begin{equation} \label{ord33}
\begin{split}
(\omega^2\rho_0+g\rho_{0z})w_{1\xi\xi}+&\frac{\omega^2}{k^2}(\rho_0w_{1z})_z  \\
& =-\frac{\omega^2}{k^2}\left(2(\rho_0u_{0X})_z
+\frac{1}{\mathrm{i}\omega}
P_{0YYz}-\frac{k_X}{k}(\rho_0u_0)_z\right)\,.
\end{split}
\end{equation}
Seeking solutions of Eq.~(\ref{ord33}) which depend on $\xi$ by the
factor $\exp(\mathrm{i}\xi)$, we obtain for $w_1$ the equation
\begin{equation} \label{ord33a}
\begin{split}
\omega^2(\rho_0w_{1z})_z -k^2(\omega^2\rho_0 + &g\rho_{0z})w_1  \\
&= -\omega^2\left(2(\rho_0u_{0X})_z+\frac{1}{\mathrm{i}\omega}
P_{0YYz}-\frac{k_X}{k}(\rho_0u_0)_z\right)\,.
\end{split}
\end{equation}
A solution to this differential equation with respect to $z$ with
the boundary conditions Eqs.~(\ref{ord25a}) and (\ref{ord26}) can be
found only when a certain compatibility condition is satisfied,
because the right hand side of Eqs.~(\ref{ord33a}) and (\ref{ord26})
contain the solution of the spectral problem (\ref{ord21}). The
compatibility condition yields the required evolution equation for
the amplitude function $A$
\begin{equation}\label{ord43}
\begin{split}
A_X+\frac{1}{2\mathrm{i}k}A_{YY}-\frac{1}{2}\frac{k_X}{k}A
&+\frac{\omega^2}{2k^2}\rho_0(-H_0)\cdot H_{0X}\cdot
\left(\phi_z(X,-H_0)\right)^2A\\
&-\frac{\omega^2}{2\mathrm{i}k}\rho_0(-H_0)\cdot H_1\cdot
\left(\phi_z(X,-H_0)\right)^2A=0\,,
\end{split}
\end{equation}
which we call the parabolic equation for periodic internal waves.
Rewritten in the initial coordinates $(x,y)$, it has the form
\begin{equation}\label{ord43prim}
\begin{split}
A_x+\frac{1}{2\mathrm{i}k}A_{yy}-\frac{1}{2}\frac{k_x}{k}A
&+\frac{\omega^2}{2k^2}\rho_0(-H_0)\cdot H_{0x}\cdot
\left(\phi_z(x,-H_0)\right)^2 A\\
&-\frac{\omega^2}{2\mathrm{i}k}\rho_0(-H_0)\cdot \bar H_1\cdot
\left(\phi_z(x,-H_0)\right)^2 A=0\,,
\end{split}
\end{equation}
where $\bar H_1(x,y)=\epsilon H_1(\epsilon x,\epsilon^{1/2}y)$,
For detailed derivation of Eq.~(\ref{ord43}) see Appendix~\ref{A2}.
\par
For the numerical calculation of the coefficients of Eq.~(\ref{ord43})
can be used, in principle, any algorithm for
solving the spectral problem Eq.~(\ref{ord21}). Some problems arise
with the derivatives, namely, $k_X$ and $\phi_z(X,-H_0))$.
\par
To avoid numerical differentiation in the calculation of $k_X$ we
differentiate the spectral problem Eq.~(\ref{ord21}) with respect to
$X$ and get the boundary value problem
 for $\phi_X$
\begin{equation*} \label{ord44}
\begin{split}
&\omega^2(\rho_0\phi_{Xz})_z-k^2(\omega^2\rho_0+g\rho_{0z})\phi_X
-2kk_X(\omega^2\rho_0+g\rho_{0z})\phi=0\,,  \\
&\phi_X(X,0)=0\,, \qquad \phi_X(X,-H_0)-H_{0X}\phi_z(X,-H_0)=0\,.
\end{split}
\end{equation*}
The compatibility condition for this problem is
\begin{equation*}\label{ord46}
\frac{k_X}{k} =  -\frac{\omega^2}{2k^2}\rho_0(-H_0)\cdot
H_{0X}\cdot\left(\phi_z(X,-H_0)\right)^2\,.
\end{equation*}
which gives the stable formula for the  numerical calculation of $k_X/k$ (modulo the calculation of
$\phi_z(X,-H_0))$).
\par
Now we derive a formula for the stable calculation of the
 derivative $\phi_z(X,-H_0)$. To do this, we multiply Eq.~(\ref{ord21})
by $z$ and integrate from $-H_0$ to zero. After integrating by parts
and some transformations we get the required formula
\begin{equation*}\label{ord47}
\phi_z(X,-H_0)=\frac{1}{\rho_0(-H_0)
H_0}\left(\frac{k^2}{\omega^2}\int^0_{-H_0}(\omega^2\rho_0+g\rho_{0z})\phi\cdot
z\,dz -\int^0_{-H_0}\rho_{0z}\phi\,dz\right)\,.
\end{equation*}

\begin{figure}[ptb]
\begin{center}
\includegraphics[width=1.0\textwidth]{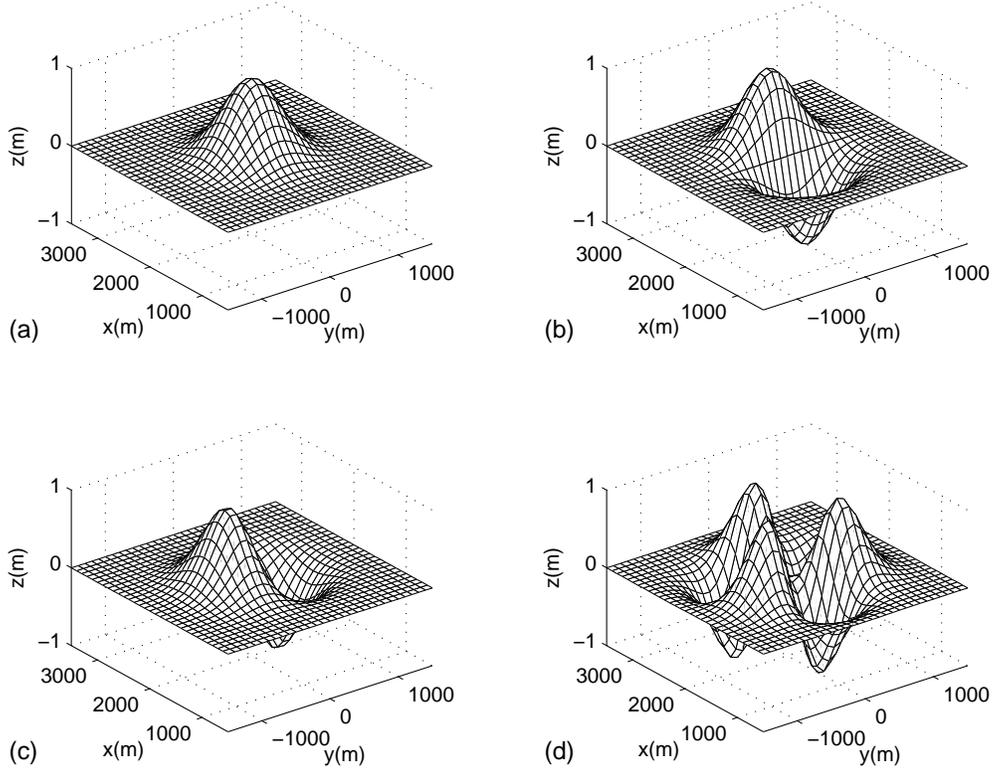}
\end{center}
\caption{Bottom topography forms of the compact inhomogeneities (\ref{coef2}) used for
the numerical simulation. For all cases $h_1=1\,\text{m}$, $\sigma=500\,\text{m}$.
(a)~$M=0$, $\alpha_0=0$;
(b)~$M=1$, $\alpha_0=0$;
(c)~$M=1$, $\alpha_0=\pi/2$;
(d)~$M=3$, $\alpha_0=0$.
}\label{r0}
\end{figure}
\begin{figure}[ptb]
\begin{center}
\includegraphics[width=5.0in]{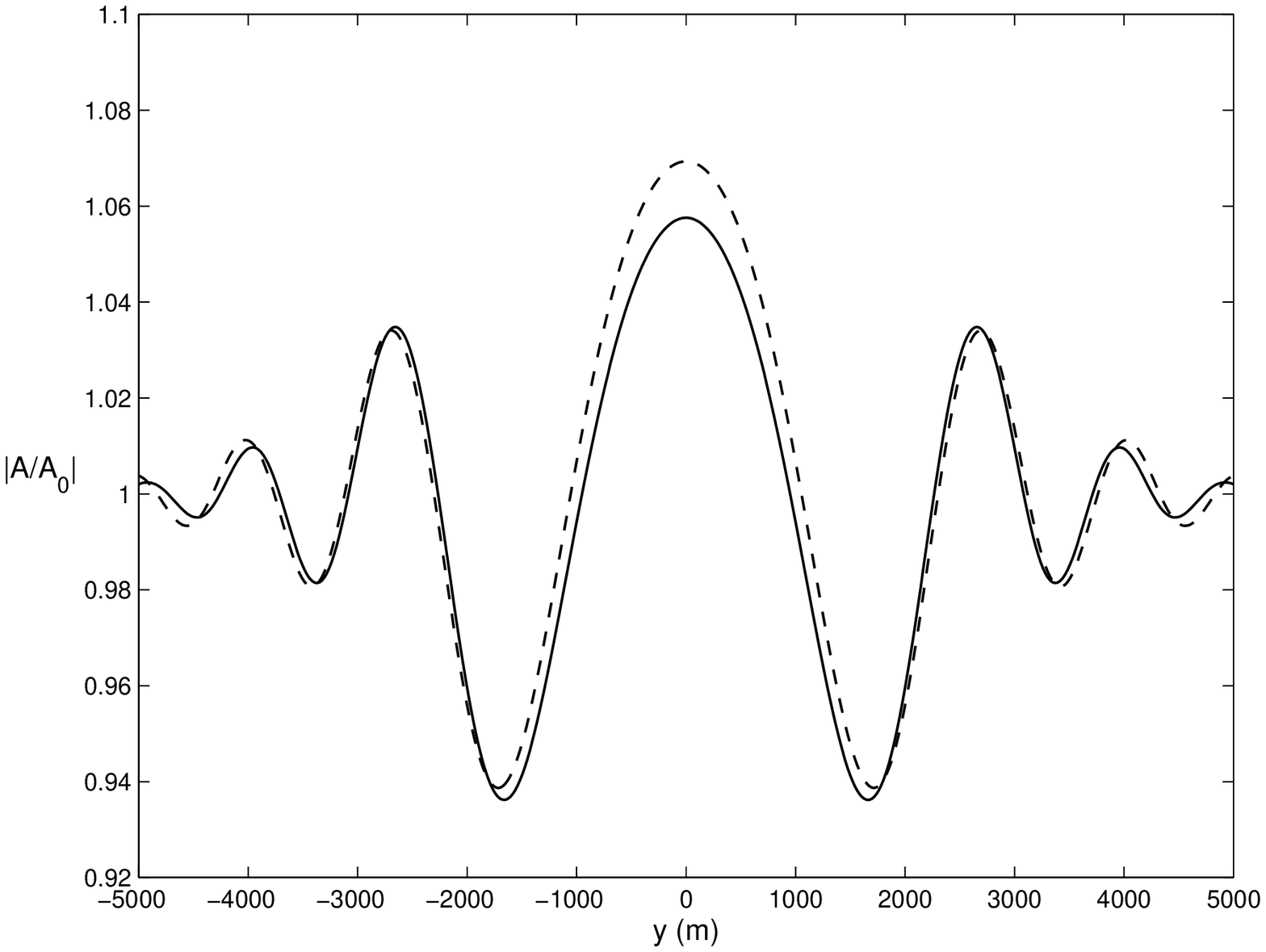}
\end{center}
\caption{Transverse cross-sections at $x=12000\,\text{m}$ of relative  wave field amplitudes in
scattering of the 2nd mode internal waves on the
shoal (\ref{coef2}) with parameters $M=0, \sigma=500~m,
\alpha_0=0$ (Fig.~\ref{r0}(a)).  ---, the parabolic equation
method; -\,-\,-\,, the Born approximation (\ref{coef3}). }\label{r1}
\end{figure}
\begin{figure}[ptb]
\begin{center}
\includegraphics[width=5.0in]{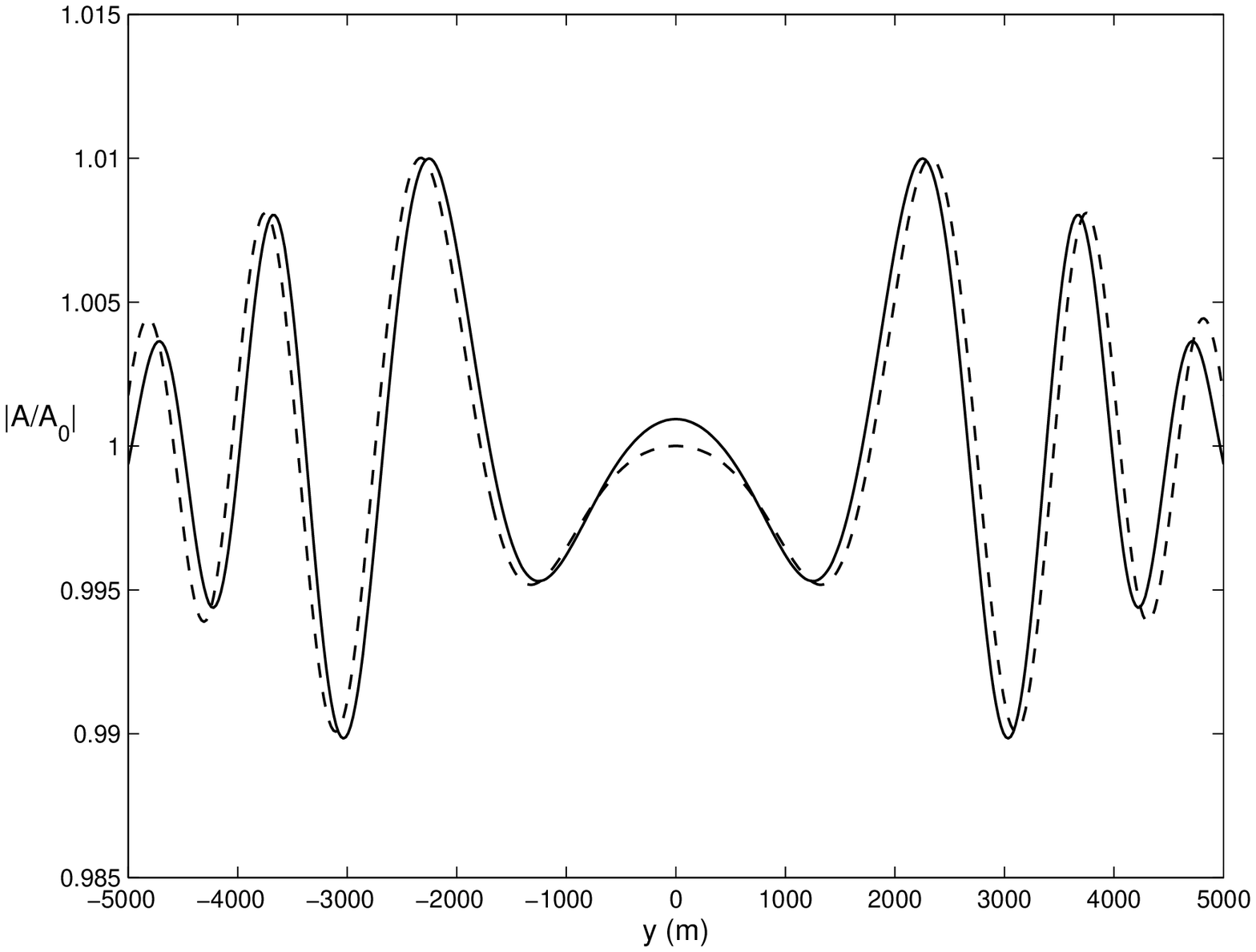}
\end{center}
\caption{Transverse cross-sections at $x=12000\,\text{m}$ of relative wave field amplitudes in
scattering of the 2nd mode internal waves on the
shoal (\ref{coef2}) with parameters $M=1, \sigma=500~m,
\alpha_0=0$ (Fig.~\ref{r0}(b)).  ---, the parabolic equation
method; -\,-\,-\,, the Born approximation (\ref{coef3}).}\label{r2}
\end{figure}
\begin{figure}[ptb]
\begin{center}
\includegraphics[width=5.0in]{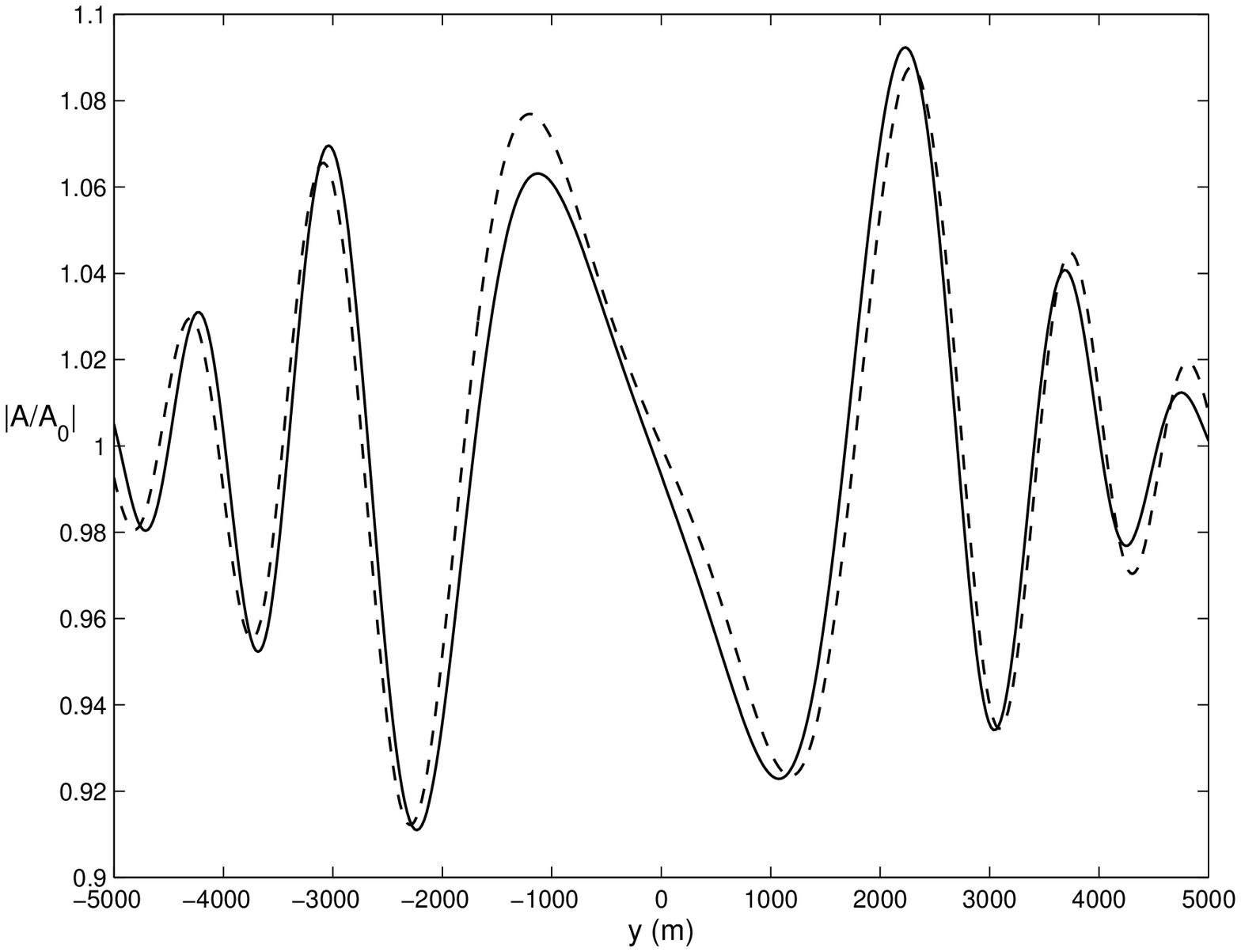}
\end{center}
\caption{Transverse cross-sections at $x=12000\,\text{m}$ of relative wave field amplitudes in
scattering of the 2nd mode internal waves on the
shoal (\ref{coef2}) with parameters $M=1, \sigma=500~m,
\alpha_0=\pi/2$ (Fig.~\ref{r0}(c)). ---, the parabolic equation method;
-\,-\,-\,, the Born approximation (\ref{coef3}).}\label{r3}
\end{figure}
\begin{figure}[ptb]
\begin{center}
\includegraphics[width=5.0in]{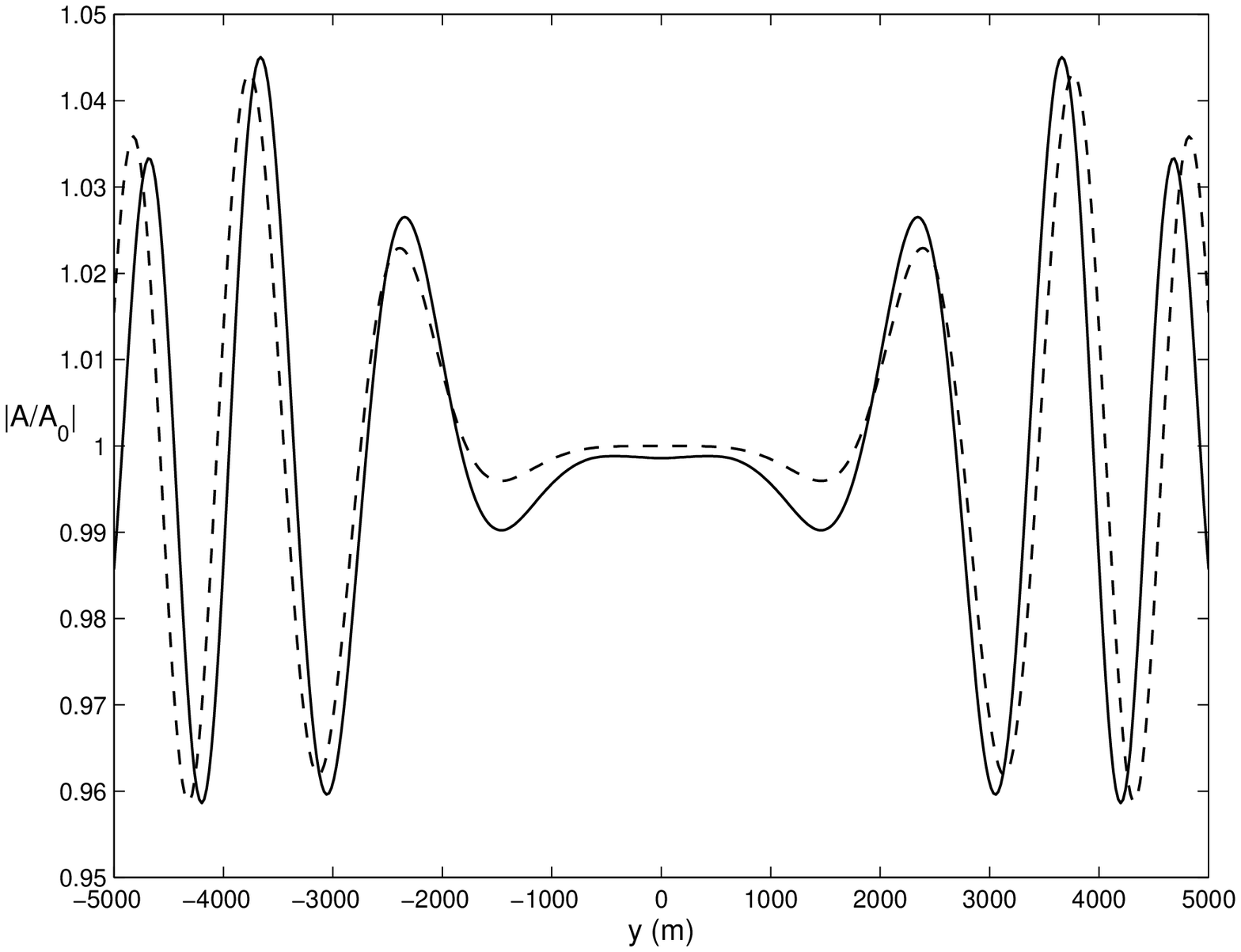}
\end{center}
\caption{Transverse cross-sections at $x=12000\,\text{m}$ of relative wave field amplitudes in
scattering of the 2nd mode internal waves on the
shoal (\ref{coef2}) with parameters $M=3, \sigma=500~m,
\alpha_0=0$ (Fig.~\ref{r0}(d)). ---, the parabolic equation
method; -\,-\,-\,, the Born approximation (\ref{coef3}).}\label{r4}
\end{figure}

\section{Numerical experiments}

The numerical experiments were conducted for a fluid with an
exponential density stratification $\rho= \exp(-\gamma z)$, where
$\gamma > 0$. In this case the complete set of normalized solutions
of the spectral problem Eq.~(\ref{ord21}) is
\begin{equation}\label{num1}
\begin{split}
k_n=\frac{\omega}{2 H_0}\sqrt{\frac{\gamma^2
H_0^2+4n^2\pi^2}{g\gamma-\omega^2}}\,,\quad n=1,\ldots\,,\\
\phi_n(z)=C\exp(\gamma z/2)\sin(n\pi z/H_0)\,,\quad n=1,\ldots\,,
\end{split}
\end{equation}
where $C$ is the normalizing constant
\begin{equation}\label{num2}
C = \sqrt{\frac{2}{H_0(g\gamma-\omega^2)}}\,.
\end{equation}
From Eqs.~(\ref{num1}) and (\ref{num2}) we have also
\begin{equation*}
\phi_{nz}(-H_0)=(-1)^nC\frac{n\pi}{H_0}\exp(-\gamma H_0/2) \,.
\end{equation*}

For the numerical experiments the problem of the scattering of
the incident plane wave of a given mode on the localized small inhomogeneities of
the bottom topography was considered. So  $H_0$ is taken to be a constant and $\bar H_1(x,y)$
describes the inhomogeneity, the total depth is $H_0+\bar H_1(x,y)$. For the inhomogeneities of the form
\begin{equation}\label{coef2}
\bar H_1(r,\alpha)=-A_Mr^M\cos(M(\alpha+\alpha_0))\exp(-r^2/\sigma^2)\,,M=1,\ldots\,,
\end{equation}
where $(r,\alpha)$ are the polar coordinates centered at the point $(x_0,y_0=0)$ with
$\alpha=0$ corresponding to the positive $x$-direction,
the scattering problem for the propagating in $x$-direction incident plane wave
of the $n$th mode
with the vertical velocity $w_{inc}=A_0\exp(\mathrm{i}k_n (x-x_0))\phi_n$ admits an approximate solution
in which the $n$th mode component of the scattered field is
\begin{equation}\label{coef3}
\begin{split}
w_n^{scat}(r,\alpha,z)=&A_n(r,\alpha)\phi_n=\left( A_0 A_M G\,
\mathrm{i}^{M+1}\,\sqrt{\frac{\pi}{2}}\,\frac{\kappa^M\sigma^{2M+2}}{2^{M+1}}
\frac{ e^{\mathrm{i}(k_nr-\pi/4)}}{\sqrt{k_nr}} \right. \\
&\left.\times \exp(-\frac{\sigma^2(k^2_n-k^2_n\cos\alpha)}{2})\cos
M(\psi-\alpha_0)\cos\alpha\right)\cdot\phi_n\,,\\
\end{split}
\end{equation}
where $G=\omega^2\rho(-H_0)\left(\phi_{nz}(-H_0)\right)^2$, $\kappa = \sqrt{2 k^2_n - 2 k^2_n\cos\alpha}$
~and $\tan(\psi)= \sin\alpha/(1-\cos\alpha)$. The quantity that will be compared with the solution of the
parabolic equation (\ref{ord43prim}) is the absolute value of the $n$th mode part amplitude of
the incident$+\,$scat\-te\-red field
$$
|A(x,y)| = |A_0\exp(\mathrm{i}k_n (x-x_0))+A_n(\sqrt{(x-x_0)^2+y^2},\arctan(y/(x-x_0)))|\,.
$$

\par
The solution Eq.~(\ref{coef3}) was obtained by the methods of the
work \cite{zakh} in the frame of the Born and far field
approximations. The methods of the work \cite{zakh} do not use any assumptions
on the preferred propagation direction and are closely related to the methods
of the work \cite{ll_smith}.

The numerical experiments were conducted on the computational domain
$[0 \le x \le 12000]\times[-5000 \le y \le 5000]$, where $x$ and $y$
are measured in meters. Eq.~(\ref{ord43prim}) with the initial
condition $A_0=A(0,y)=constant$ was integrated on the grid $395\times349$ by
the Crank-Nicholson scheme \cite{pot}. In order to exclude
reflections at the boundaries, the absorbing Baskakov-Popov boundary
conditions were used~\cite{baspop}, which were adapted for
non-vanishing at the boundaries initial conditions.

The inhomogeneities in all cases have width parameter $\sigma =
500\,\mbox{m}$ and the amplitude $h_1=A_M\exp(-M/2)\left(\sigma\sqrt{M/2}\,\right)^M=1\,\text{m}$
 with the center positioned
at $x_0 = 2000\,\mbox{m}$, $y_0 = 0\,\mbox{m}$.
The shape parameters $M$ and $\alpha_0$ were varied (see Fig.~\ref{r0}).
The depth of the flat bottom, $H_0$, was taken to be 60~m. The
density stratification parameter $\gamma=0.00025$, which corresponds
to the Brunt-V\"ais\"al\"a frequency $N=\sqrt{g\gamma}\approx
0.05\,s^{-1}$.

The computations were done for the  1st mode and 2nd mode incident
plane waves with 15~min time period. The transverse cross-sections of the 2nd mode
computed field amplitude at $x=12000~\text{m}$ are presented in
Fig.~\ref{r1}-\ref{r4} in comparison with the Born type
approximation scattering results obtained by Eq.~(\ref{coef3}).
The results for the 1st mode are analogous.
\par
Considering the results of computations, it is worth noting that the solution (\ref{coef3})
has an approximative character, so the comparison with it is not exactly the test of accuracy
of the derived parabolic equation. Nevertheless, since Eq.~(\ref{coef3}) is of quite different genesis,
and, in particular, free from any assumptions on the preferred propagation direction, this comparison can
lead to the conclusion that the parabolic equation describes sufficiently well the waves with propagation
angles up to $45^{\,\circ}$ (see the discussion on the propagation angles in \cite{dal-kir}),
scattering on the enough rough topography.

\section{Conclusion}

For the propagation of periodic internal waves over uneven bottom
topography with small irregularities, the  narrow-angle parabolic
equation (\ref{ord43prim}) has been derived. It also takes into account
slow, but not necessary small, variations of bottom topography
 in principal propagation direction.
\par
%The results of numerical simulations support the validity of
%equation (\ref{ord43}) in the scope specified during its derivation.
We have illustrated the use of the obtained equation by presenting
the results of scattering of the plane internal waves over  shoals
of the special forms~(\ref{coef2}). The results of computations are in
a sufficiently good agreement with the analytical solution~(\ref{coef3}),
obtained in the frame of the Born approximation,
and support the applicability of equation (\ref{ord43prim}) for computing
of internal wave fields over uneven bottom with restrictions typical for
the parabolic equation method in general \cite{tap,mei}.

\appendix
\section{Derivation of Eq.~(\ref{ord33})}\label{A1}
From  (\ref{ord23}, \ref{ord25}) we obtain
\begin{equation}\label{ord28}
(\omega^2\rho_0+g\rho_{0z})w_1+\mathrm{i}\omega P_{1z}=0\,.
\end{equation}
From (\ref{ord22}) we get
\begin{equation}\label{ord29}
P_{1\xi\xi}=\frac{1}{\Theta_X}\mathrm{i}\omega\rho_0u_{1\xi}-
\frac{1}{\Theta_X}P_{0X\xi}\,.
\end{equation}
From (\ref{ord24}), taking into account (\ref{ord27}), we have
\begin{equation}\label{ord30}
u_{1\xi}=-\frac{1}{\Theta_X}(u_{0X}+v_{1/2Y}+w_{1z})=-\frac{1}{\Theta_X}
(u_{0X}+\frac{1}{\mathrm{i}\omega\rho_0}P_{0YY}+w_{1z})\,.
\end{equation}
From (\ref{ord13}) we get
\begin{equation}\label{ord31}
P_{0X\xi}=\mathrm{i}\omega\left(\frac{1}{\Theta_X}\rho_0u_0\right)_X
=-\mathrm{i}\omega\frac{k_X}{k^2}\rho_0u_0
+\mathrm{i}\omega\frac{1}{k}\rho_0u_{0X}\,.
\end{equation}
Substitution of $u_{1\xi}$ and $P_{0X\xi}$ from Eqs.~(\ref{ord30}, \ref{ord31}) into  Eq.~(\ref{ord29})
gives
\begin{equation}\label{ord32}
\begin{split}
P_{1\xi\xi}&=-\frac{1}{k^2}\mathrm{i}\omega\rho_0\left(u_{0X}+\frac{1}{\mathrm{i}\omega\rho_0}P_{0YY}+
w_{1z}\right)\\
&\qquad \qquad\qquad\qquad\qquad\quad
-\frac{1}{k^2}\left(-\mathrm{i}\omega\frac{k_X}{k}\rho_0u_0+
\mathrm{i}\omega\rho_0u_{0X}\right)\\
&=-\frac{1}{k^2}\mathrm{i}\omega\left(2\rho_0u_{0X}+
\frac{1}{\mathrm{i}\omega}P_{0YY}+\rho_0w_{1z}-\frac{k_X}{k}\rho_0u_0\right)\,.
\end{split}
\end{equation}
Differentiating Eq.~(\ref{ord32}) with respect to $z$ and
substituting the result into the second $\xi$-derivative of
 Eq.~(\ref{ord28}), we obtain Eq.~(\ref{ord33}).

\section{Derivation of the compatibility condition}\label{A2}

In this appendix the compatibility condition for the boundary value
problem Eqs.~(\ref{ord33a}, \ref{ord25a} and \ref{ord26}) is
derived.
\par
From Eq.~(\ref{ord18}) we have
\begin{equation*} \label{ord34}
\frac{1}{\mathrm{i}\omega}P_{0YYz}=\frac{1}{\omega^2}(\omega^2\rho_0+g\rho_{0z})w_{0YY}\,.
\end{equation*}
From Eq.~(\ref{ord16}), taking into account that $u_0$ and $w_0$
depend on $\xi$ by the factor $e^{\mathrm{i}\xi}$, we have
\begin{equation*} \label{ord35}
\mathrm{i}u_0=-\frac{1}{k}w_{0z},\quad \mbox{or}\quad
u_0=\frac{\mathrm{i}}{k}w_{0z}\,.
\end{equation*}
Substitution of these expressions into the right hand side of~(\ref{ord33a})
(denote it by $RHS$) yields
\begin{equation} \label{ord36}
\begin{split}
RHS= &-\omega^2\cdot\left[ \vphantom{\frac{\mathrm{i}k_X}{k^2}}
2\frac{\mathrm{i}}{k}(\rho_0w_{0zX})_z-2\frac{\mathrm{i}}{k^2}k_X(\rho_0w_{0z})_z
 +\frac{1}{\omega^2}\left(\omega^2\rho_0+g\rho_{0z}\right)w_{0YY}\right.\\
&-\left.\frac{\mathrm{i}k_X}{k^2}(\rho_0w_{0z})_z\right]=
 -\omega^2\cdot\left[ \vphantom{\frac{\mathrm{i}k_X}{k^2}}2\frac{\mathrm{i}}{k}A_X(\rho_0\phi_z)_z +
 2\frac{\mathrm{i}}{k}A(\rho_0\phi_{Xz})_z\right.\\
&\left.
+\frac{1}{\omega^2}\left(\omega^2\rho_0+g\rho_{0z}\right)A_{YY}\phi
-
3\frac{\mathrm{i}k_X}{k^2}A(\rho_0\phi_z)_z
\right]\cdot\exp(\mathrm{i}\xi)\,.
\end{split}
\end{equation}
To obtain the required compatibility condition we multiply
Eq.~(\ref{ord33a}) by the eigenfunction $\phi$ and integrate with
respect to $z$ from $-H_0$ to $0$. Then twice integrating by parts Eq.~(\ref{ord33a})
using Eqs.~(\ref{ord25a}, \ref{ord26}) we obtain
\begin{equation} \label{ord37}
\int^0_{-H_0}RHS\cdot\phi\,dz=\omega^2\rho_0(-H_0) a\phi_z(X,-H_0)\,,
\end{equation}
where
\begin{equation*} \label{ord38}
\begin{split}
a=&w_1(X,-H_0)=w_{0z}(X,-H_0)\cdot H_1-u_0(X,-H_0)\cdot H_{0X}  \\
&=(H_1-\frac{\mathrm{i}}{k}H_{0X})w_{0z}(X,-H_0)=
(H_1-\frac{\mathrm{i}}{k}H_{0X})A\phi_z(X,-H_0)\cdot\exp(\mathrm{i}\xi)\,.
\end{split}
\end{equation*}
From  Eq.~(\ref{ord36}) and Eq.~(\ref{ord37}) we get
\begin{equation} \label{ord40}
\begin{split}
&2\frac{\mathrm{i}}{k}A_X\int^0_{-H_0}(\rho_{0}\phi_z)_z\phi\,dz
+ \frac{1}{\omega^2}A_{YY}\int^0_{-H_0}(\omega^2\rho_0+g\rho_{0z})\phi^2dz\\
&-3\mathrm{i}\frac{k_X}{k^2}A\int^0_{-H_0}(\rho_0\phi_z)_z\phi dz+
\frac{\mathrm{i}}{k}A\int^0_{-H_0}2(\rho_0\phi_{Xz})_z\phi\,dz\\
 &\qquad \qquad \qquad \qquad \quad=-\rho_0(-H_0)
 \cdot (H_1-\frac{\mathrm{i}}{k}H_{0X})A\cdot \left(\phi_z(X,-H_0)\right)^2\,.
\end{split}
\end{equation}
Using the normalizing condition Eq.~(\ref{ord39}) we have
\begin{equation*} \label{ord41}
\begin{split}
&\left(\frac{\omega^2}{k^2}\int^0_{-H_0}\rho_{0}(\phi_z)^2\,dz\right)_X\\
&= -2\frac{k_X}{k}
 +\frac{\omega^2}{k^2}\int^0_{-H_0}2\rho_0\phi_z\phi_{Xz}\,dz-
\rho_0(-H_0)\cdot\left(\phi_z(X,-H_0)\right)^2H_{0X}\frac{\omega^2}{k^2}=0\,,
\end{split}
\end{equation*}
and thus
\begin{equation*} \label{ord42}
\begin{split}
\frac{\mathrm{i}}{k}A\int^0_{-H_0}2(\rho_0\phi_{Xz})_z\phi
\,dz
&=-\frac{\mathrm{i}}{k}A\int^0_{-H_0}2\rho_0\phi_{Xz}\phi_z\,
dz\\
&
 =-\frac{\mathrm{i}}{k}A\rho_0(-H_0)\cdot\left(\phi_z(X,-H_0)\right)^2 H_{0X}-2\mathrm{i}\frac{k_X}{\omega^2}A\,.
\end{split}
\end{equation*}
With this and the normalizing condition Eq.~(\ref{ord39}) we obtain from Eq.~(\ref{ord40})
 the required parabolic equation
\begin{equation*}
\begin{split}
A_X+\frac{1}{2\mathrm{i}k}A_{YY}-\frac{1}{2}\frac{k_X}{k}A
&+\frac{\omega^2}{2k^2}\rho_0(-H_0)\cdot H_{0X}\cdot
\left(\phi_z(X,-H_0)\right)^2A\\
&-\frac{\omega^2}{2\mathrm{i}k}\rho_0(-H_0)\cdot H_1\cdot
\left(\phi_z(X,-H_0)\right)^2A=0\,.
\end{split}
\end{equation*}

\end{document}